# A multidimensional approach for context-aware recommendation in mobile commerce

Maryam Hosseini-Pozveh
Department of Computer Engineering
University Of Isfahan
Isfahan, Iran
m.hosseini@eng.ui.ac.ir

Mohamadali Nematbakhsh
Department of Computer Engineering
University Of Isfahan
Isfahan, Iran
nematbakhsh@eng.ui.ac.ir

Naser Movahhedinia
Department of Computer Engineering
University Of Isfahan
Isfahan, Iran
naserm@eng.ui.ac.ir

*Abstract*—**Context as the dynamic information describing the situation of items and users and affecting the user's decision process is essential to be used by recommender systems in mobile commerce to guarantee the quality of recommendation. This paper proposes a novel multidimensional approach for context-aware recommendation in mobile commerce. The approach represents users, items, context information and the relationship between them in a multidimensional space. It then determines the usage patterns of each user under different contextual situations and creates a new 2-dimensional recommendation space and does the final recommendation in that space. This paper also represents an evaluation process by implementing the proposed approach in a restaurant food recommendation system considering day, time, weather and companion as the contextual information and comparing the approach with the traditional 2-dimensional one. The results of comparison illustrates that the multidimensional approach increases the recommendation quality.**

## I. INTRODUCTION

Recommender systems in mobile commerce have been one of the important research issues in recent years which with advances in mobile and wireless computing have been concerned. Mobile or pervasive commerce is known as doing e-commerce activities via wireless environment, especially wireless internet, and mobile handled devices which are growing very fast [1], [2]. Mobile commerce applications have two special properties, mobility and broad reach [1], [3]. The former emphasizes breaking the not-anywhere limitations and the latter emphasizes breaking the not-any time limitations in the interactions between the users and the applications [1], [3], [4], [5]. The opportunity that users can simply use mobile phones or personal digital assistance at any place and any time for doing the electronic commerce activities such as e-banking, e-shopping and many other opportunities for them and also for businesses make a clear insight to consider mobile commerce applications as important research topics [1], [3]. Implementation of recommender systems in mobile commerce could not be very useful without considering the effect of unique parameters of that environment which are named context information [6].

The goal of recommendation systems is to propose the suitable resources means items such as web pages, books or movies users prefer to other items [7], [8], [9]. In recommendation systems, three data sets, user's information (C), recommendable items information (S) such as books, movies, music and so on, and the relationship data between users and items exist. The relationship between S and C is based on a rating structure that describes the usefulness degree of items to users. This relationship could be defined with a function is named utility function, u: [7]

$$u : C \times S \rightarrow Ratings \qquad (1)$$

Where Ratings is a totally ordered set including nonnegative integers or real numbers within a certain range.

The main problem of recommendation systems is that u function has defined only on a subset of $C \times S$ domain, not on the whole set and therefore unspecified parts of it must be predicted. After the prediction phase, system could recommend the items with highest predicted rates to the users [7].

To reaching the final goal in recommendation systems, different methods are developed so far which are categorized as follow: [7], [9], [10]

- Content-based: in this group of methods, system recommends items that are most similar to the items which user has rated them highly in the past. It means u(c, s) is estimated based on the utilities $u(c, s_i)$ which $s_i$ are items that are similar to s.
- Collaborative filtering: in this group of methods, system recommends items which are highly rated by peers of current user. More formally, the utility u(c, s) is estimated based on utilities $u(c_j, s)$ that users $c_j$ are users who are similar to user c.

- Hybrid models: these approaches combine two previous methods and therefore use the benefits of both of them for specifying and recommending suitable items.

From the other point of view, recommendation approaches, content-based and collaborative filtering, are divided into model-based and memory-based approaches. Opposite of memory-based models, model-based approaches make a model using rating sets and machine learning methods and use that model for future rating predictions [7], [10], [11].

The rest of this paper is organized as follows. Section 2 gives an overview of related works. Section 3 describes context-aware recommendation concepts in mobile commerce. Section 4 details the novel multidimensional approach. Section 5 presents the implementation experiences and reports the evaluation results and finally the paper concludes in section 6.

## II. RELATED WORK

Mobile handled devices and wireless technologies emergence has made many opportunities for electronic commerce applications. Presenting more customized and personalized information is one of the most important goals of the mobile commerce applications. Using context as the dynamic information describing the situation of items and users and affecting the user's decision process, by recommender systems in mobile commerce is a solution to guarantee the quality of recommendation. All the proposed methods so far, have tried to use contextual information for producing proper outputs.

Location-aware recommender systems are an important subset of context-aware ones. Yang, Cheng and Dia [12] have presented a location-aware recommender system in mobile environments that its goal is to recommend vendors websites considering customer preferences and also his/her location distance from the location presented in websites. Proximo [25] is another location-aware recommender system for indoors environments such as museums and art galleries. It shows the recommendable items on a map in the user's mobile device.

In addition to the location parameter, using other contextual parameters for recommendation has been concerned by researchers. Li, Wang, Geng and Dai [13] have proposed a context-aware recommender system for mobile commerce applications. That framework uses multidimensional model for representing recommendation space and a reduction-based approach for reducing that space to a 2-dimensional one. It then uses a traditional recommendation method in that final space.

Some mobile context-aware recommender systems use ontology and semantic web [26], [27]. Ontology could be used for modeling context or modeling relation between context and other data sets. It could also be used in recommendation process.

## III. MOBILE CONTEXT-AWARE RECOMMENDER SYSTEMS

Context-awareness concept has been used in various researches belonging to different fields of mobile computing scope and defined in them as a general definition or a special one conforming to the application usage. As Doulkeridis, Loutas and Vazirgiannis [14] present, Context could be defined as facts which include other things and add meaning to them. In mobile computing, the first reasoning of those facts is awareness of user's situation and environment surrounding her/him. Dey and Abowed [15] define context as any information which helps describing the situation of an entity. In this definition, entity is known as a person, place, or any other thing that relates to the interactions between user and application including user and application themselves. Many other articles in the context scope use the latter definition in their researches [16], [17], [18].

In the most of those mentioned researches, context has been defined as some parametric examples in addition to the general conceptual definitions. In general, context can be defined as a set of user's personal information, his/her preferences and interests, his/her current activities and environmental parameters surrounding him/her including geographical information such as location, direction and speed, environmental information such as temperature and weather, time and social parameters [6], [14], [16], [17], [19]. Dey and Abowed [15] have proposed a two-tiered architecture to define and categorize contextual parameters. At the first level, primary context types which are location, time, identity and activity are located. Other contextual parameters, secondary context types, are located at the second level and are considered as the attributes of the first level types. For example, weather or temperature can be retrieved from the location and time information.

Considering context as an input to information providing phase causes to more effective results [14], [19]. Context-aware applications adapt information/service providing with the context changes [14], [17] and thus bring to the better use of information/services for the users in different situations around them [6], [16], [19].

In the current research, the main goal is to propose an approach for recommendation in mobile commerce considering context information. Context could be used to filter or prioritize services/information for the users [6], [16], [19] and therefore using it in recommendation techniques may be very beneficial. In fact, context describes the users' dynamic properties and system can inference users' needs from the context content [16].

A mobile context-aware recommendation system must be able to recommend the items considering items/users' dynamic properties which describe the situation of users and items and affect the users' decision process.

Context information in mobile recommendation systems can be divided as follows:

1- Historical (offline) contextual information

The knowledge of which are the best users' favorite items/services in different context scopes and over time could be registered in the system for using in future estimations. Using those parameters helps to better compute the similarity degree among users and items.

2- Online contextual information

Online contextual information includes users (and items/services) in the time of sending the request to the system. A context-aware recommendation system uses this kind of information in a filtering mechanism or prioritizing process for delivering information/services to users.

In the context parameters (historical or online) definition phase, it is necessary to determine the effectiveness degree of them on making the suitable output for the system. This issue is an important sub-problem in context-aware recommendation systems. Therefore context-aware recommendation systems may include both historical and online parameters or only use online parameters to recommend items. Suppose that system only uses online context, so if user is in $context_i$, the system could recommend items which are in $context_i$ too but if system uses both online and historical context, it could recommend items which are predicted to be liked by the user in $context_i$ not all the items existing in that context.

When a system uses contextual information, it must be modeled and the relationship between it and other system information sets of the system must be determined. Adomavicius, Sankaranarayanan, Sen and Tuzhilin [4] have proposed a multidimensional model for context-aware recommender systems that can be used in mobile recommender systems too. Therefore the historical contextual information could be represented in multidimensional model and recommendation space changes from a 2-dimensional space to a multidimensional one that each of the historical parameters is one of its dimensions. Presenting each dimension as Di, utility function is as below: [10]

$$u: D_1 \times D_2 \times_{...} D_n \rightarrow \text{Ratings} \quad (2)$$

Where each $D_i$ (each of dimensions) is a subset of Cartesian product of its attributes and each attribute is a set of values:

$$D_i \subseteq A_{ik1} \times A_{ik2} \times \ldots \times A_{iki} \quad (3)$$

Where, each dimension is supposed to have $K_i$ attribute. These attributes are called profile of the dimension (for example user dimension includes name, sex, age and so on).

If a mobile recommender system doesn't use historical contextual information, multidimensional model changes to a classical 2-dimensional one. Online contextual parameters are delivered to system with users' request. Online parameters set may not exactly equal with the historical parameters set. In fact set of historical parameters could be a smaller or equal subset of online parameters set.

## IV. THE MULTIDIMENSIONAL APPROACH FOR MOBILE CONTEXT-AWARE RECOMMENDER SYSTEMS

Context information affects user's decision process. In a recommendation system, it means that the item sets which a user likes could be different in various context situations, $context_1$, $context_2$ to $context_n$. Therefore, the proposed approach uses a multidimensional dataset with the cube of ratings and it includes phases as follow:

(1) Recognizing Users' different usage patterns under different context situations: As Ehrig, Hasse, Hefke and Stojanovic define [20]"If in two contexts the same (related) entities are used then these contexts are similar", the context situations that a user has similar usage patterns in them would be determined. In fact, different context situations that have been defined for the system would be clustered based on the user's usage patterns. The clusters would be labeled from 1 to m (m is the number of clusters).

(2) Making a new 2-dimensional recommendation space: In this phase, using the results of the previous phase, for each user $c_i$, m new user $c_{i1}$ to $c_{im}$ would be defined. m could be the same or not for different users depending on the method that is used in phase one. In fact, if $c_i$ in $context_s$, $context_g$ and $context_c$ has similar usage patterns and those contexts are labeled with $context_{out\text{-}L}$ (1<L<m), as it is shown in "Fig. 1" a new user $c_{i\text{-}L}$ would be defined. User $c_{i\text{-}L}$ is equivalent with user $c_i$ in different context situations labeled with $context_{out\text{-}L}$.

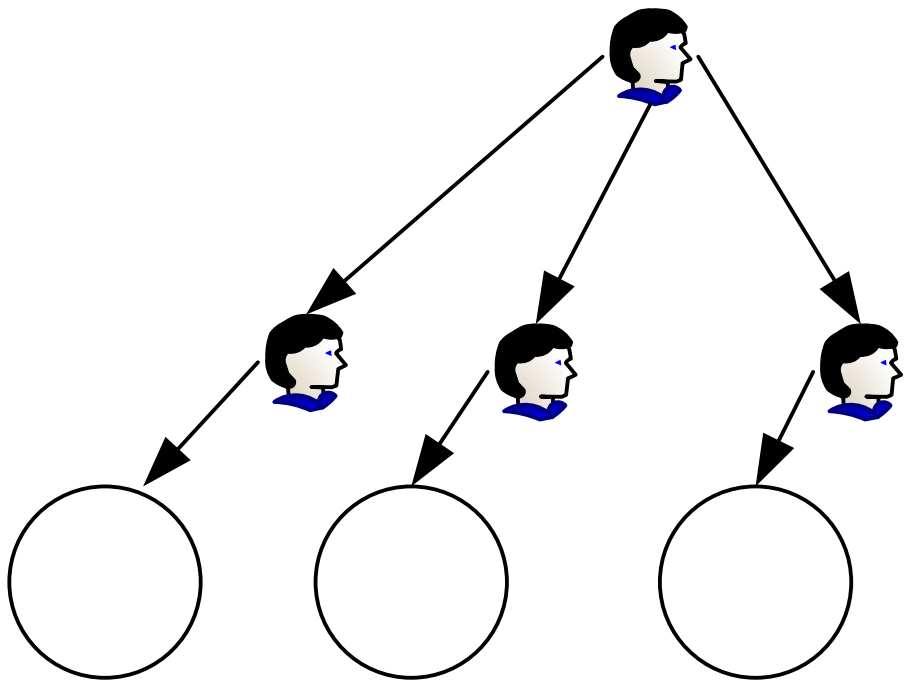

Figure 1. Making new users equivalent with user $c_i$ in different contextual situations

So the relation (2) could be redefined as relation (4):

$$u : C' \times S \rightarrow \text{Ratings}' \quad (4)$$

Where $C'$ is the new user set, S is the same as in relation (2) and $\text{Ratings}'$ is the new users' rating set. If $c_i$ has rated item $s_i$ in different context situations labeled with $context_{out\text{-}L}$(1<L<m), the user $c_{i\text{-}L}$'s rate to item $s_i$ in relation(4) is the result of a aggregation function (such as average) on the previous ratings.

(3) Doing the recommendation process in the new 2-dimensional space: Any of the model-based or memory-based approaches then could be used in this phase to recommend suitable items to users.

## V. IMPLEMENTATION AND EVALUATION

For evaluating the proposed approach, a mobile context-aware recommender system for restaurant food items is used. First, ratings are gathered and then the method is evaluated on that data set. Various dimensions of the system are user and item as the main dimensions and day, time, weather and companion as the contextual dimensions:

- Day: Weekday, Weekend.
- Time: Morning, Noon, Afternoon, Night.
- Companion: Spouse, Family, Friends, Co-workers, Alone, Others.
- Whether: Cold/Sunny, Cold/Rainy, Moderate/Sunny, Moderate/Rainy, Hot/Sunny, Hot/Rainy, others.

In this research, location is not used as a historical parameter, because it is not differed between the same food items of various restaurants in the recommendation scenario.

Data set includes ratings of 630 users to 400 foods in different contextual situations from October to December 2008. The system used for gathering the ratings is implemented with J2ME technology (see http://java.sun.com/ for its description). Evaluation is done offline and with division of the set to 80% (training set) and 20% (test set).

For the usage pattern recognition and for the similar users recognition, self-organizing maps are used. Self-organizing maps (i.e. SOMs) or kohonen model are a kind of unsupervised artificial neural networks. Because they could be used as a clustering technique, they are used in many recommendation systems, especially in collaborative filtering-based recommendation systems. They have been proved to have good performance in the recommendation quality and processing costs [11] [21]. As a simplified definition for them, in a topology-preserving map, units (neurons) located physically next to each other will respond to classes of input vectors that are likewise next to each other. It means these networks find the similar input vectors or cluster them [22], [23] ("Fig. 2").

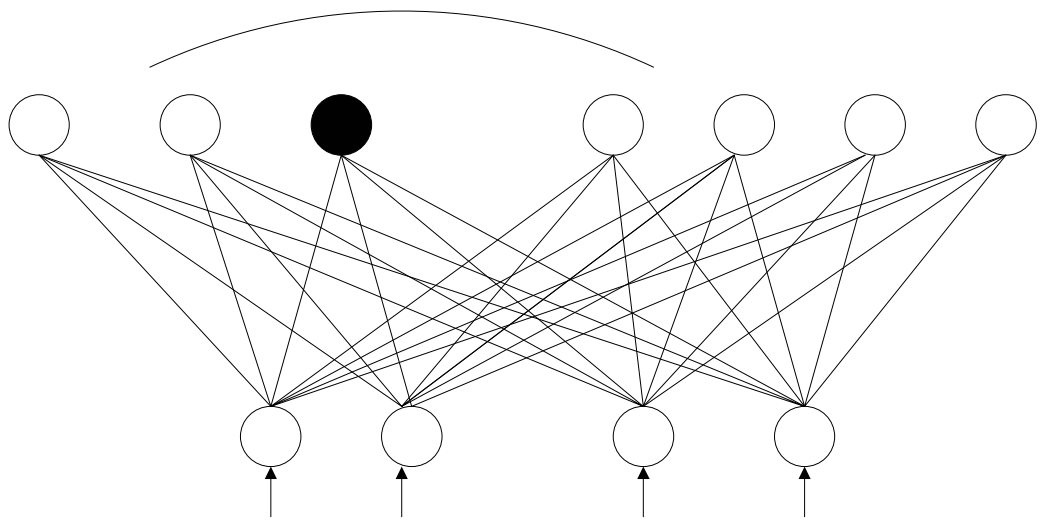

Figure 2. Self-organizing map

For each input vector fired node and its neighborhood nodes are updated as below [22], [23]:

$$\mathrm{W}(\mathrm{t}+1)=\mathrm{W}(\mathrm{t})+\alpha(X-W(t)) \qquad (5)$$

In the current paper, cosine similarity measure is used to compute the distance between input vectors and weight vectors of units:

$$\cos(\vec{x},\vec{w})=\frac{\vec{x}.\vec{w}}{\|\vec{x}\|_2\times\|\vec{w}\|_2}=\frac{\sum_{s\in S} r_{x,s}r_{w,s}}{\sqrt{\sum_{s\in S} r_{x,s}^2\sum_{s\in S} r_{w,s}^2}} \qquad (6)$$

In the usage pattern recognition phase, different contextual situations are clustered for each user. So each input vector of the self-organizing map is one of the context situations (number of the inputs is equal with the production of the number of contextual dimensions' values). The vectors are defined in a p-dimensional space, where p is the number of the items. Values of the vectors are user's rating to the items. After user's usage pattern recognition phase, the new recommendation space is built. In the next phase, using SOM, a collaborative filtering-based recommendation method is used. Self-organizing map networks is used for clustering the users. So each input vector is one of the users of the system. The vectors are defined in a p-dimensional space, where p is the number of items. Values of the vectors are user's rating to the items.

With changing the node numbers of kohonen layer and considering its effect on the F1 measure [24], the optimal numbers of the neurons for the usage pattern recognition network and similar users' recognition network have been determined. In the former network, they were changed from 2 to 15 and in the later they were changed from 5 to 35. The final neuron numbers were determined to be 6 neurons (resulted in 6 clusters) in the former network and 21 neurons (resulted in 21 clusters) in the latter network. F1 measure is defined as below [24]:

$$\mathrm{F1}=\frac{2.\mathrm{Precision}.\mathrm{Recall}}{\mathrm{Precision}+\mathrm{Recall}}. \qquad (7)$$

Where Precision is defined as the ratio of relevant items selected to number of items selected and Recall is defined as the ratio of relevant items selected to total number of relevant items available.

Average F1 for 200 users and Top-5, Top-10, Top-15, Top-20, Top25 and Top-30 is calculated. The results are presented in "Fig. 3".

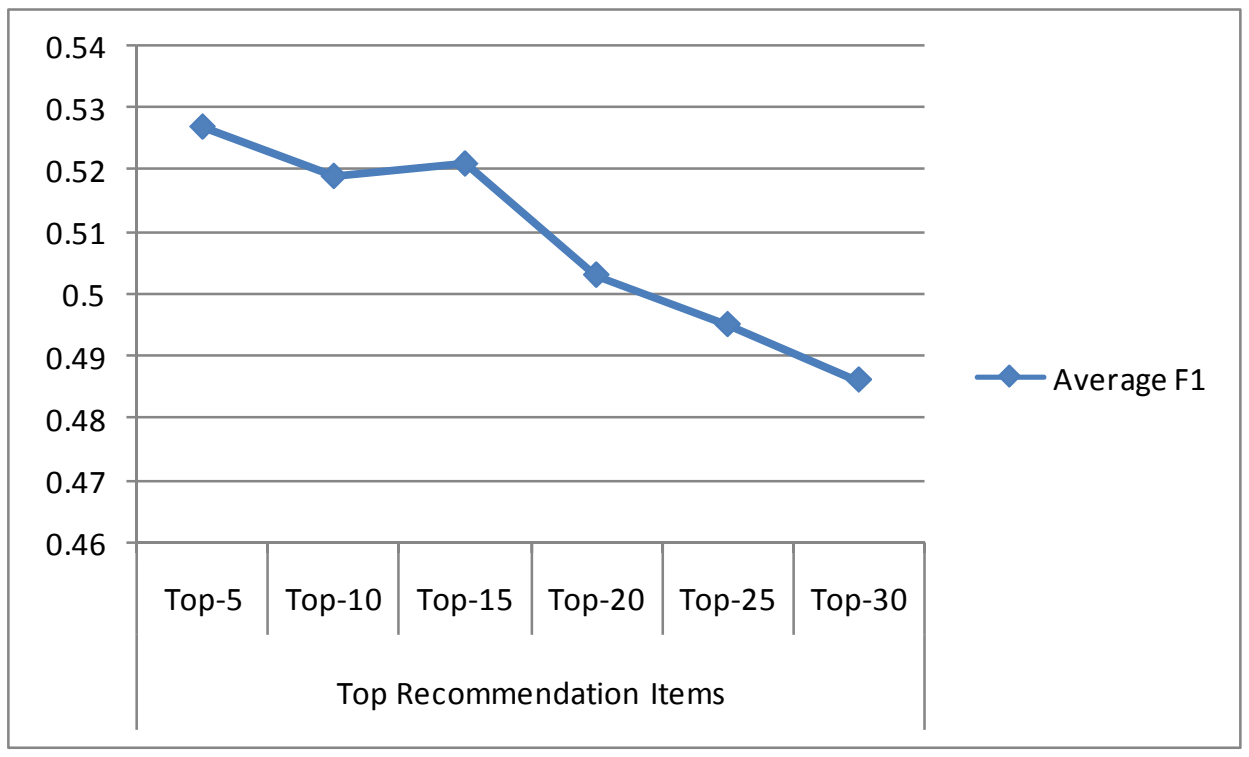

Figure 3. Average F1 measure diagram for the multidimensional recommendation approach

The proposed approach is compared with the traditional recommendation approach without considering the contextual parameters in recommendation. The gathered dataset is used in this phase too. Some items are rated only in one context situation but some items are rated in more than one context situation by a single user. In the former, the values of user and item dimensions and the rating corresponding to them are selected. In the latter, the values of user and item dimensions are selected and the rating corresponding to them is calculated with a aggregation function (for example with the average function) between all those ratings.

This method is also implemented with SOMs as a collaborative filtering-based way in the 2-dimensional space. The optimal numbers of the neurons in the kohonen layer was determined to be 19 (resulted in 19 clusters). Average F1 measure for 200 users and Top-5, Top-10, Top-15, Top-20, Top25 and Top-30 is calculated. As the results which are presented in "Fig. 4" show, the recommendation quality is increased in context-aware multidimensional recommendation approach.

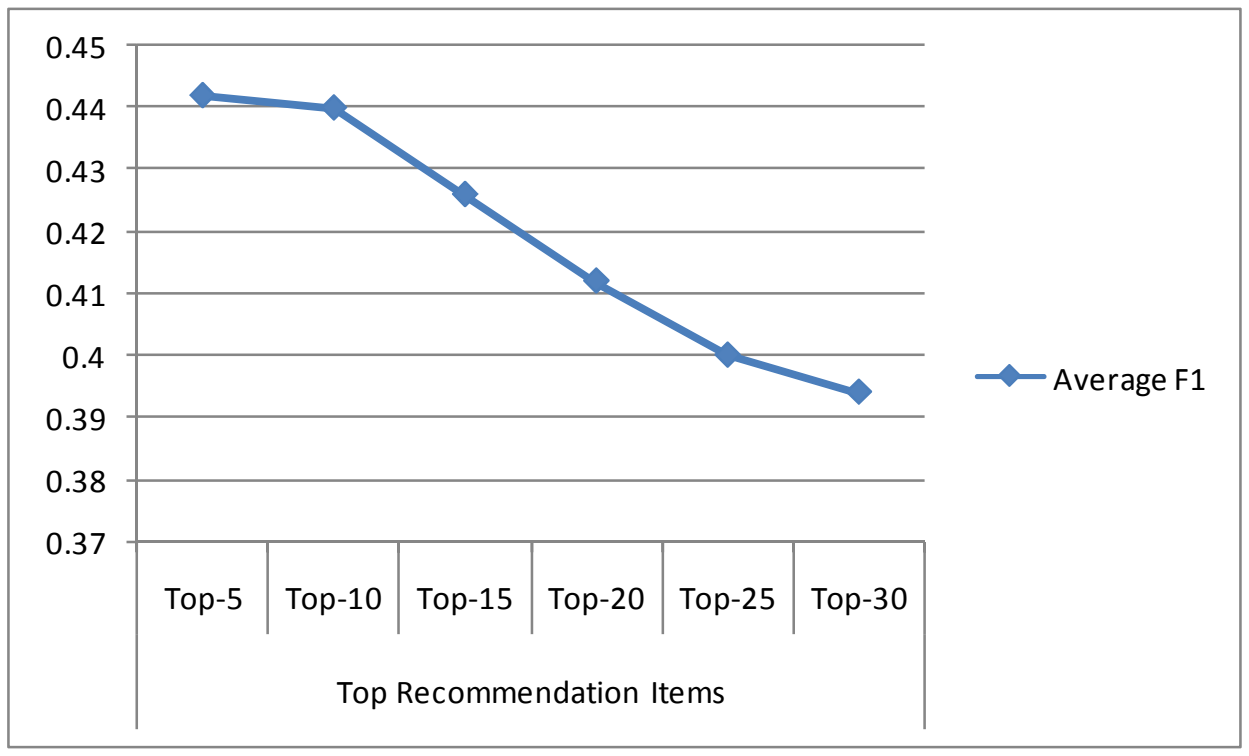

Figure 4. Average F1 measure diagram for the traditional recommendation approach

Average F1 measure is also calculated for 10 user of each cluster and for Top-5 recommendation items. The result for both multidimensional approach and traditional approach are presented in "Fig. 5" and "Fig. 6".

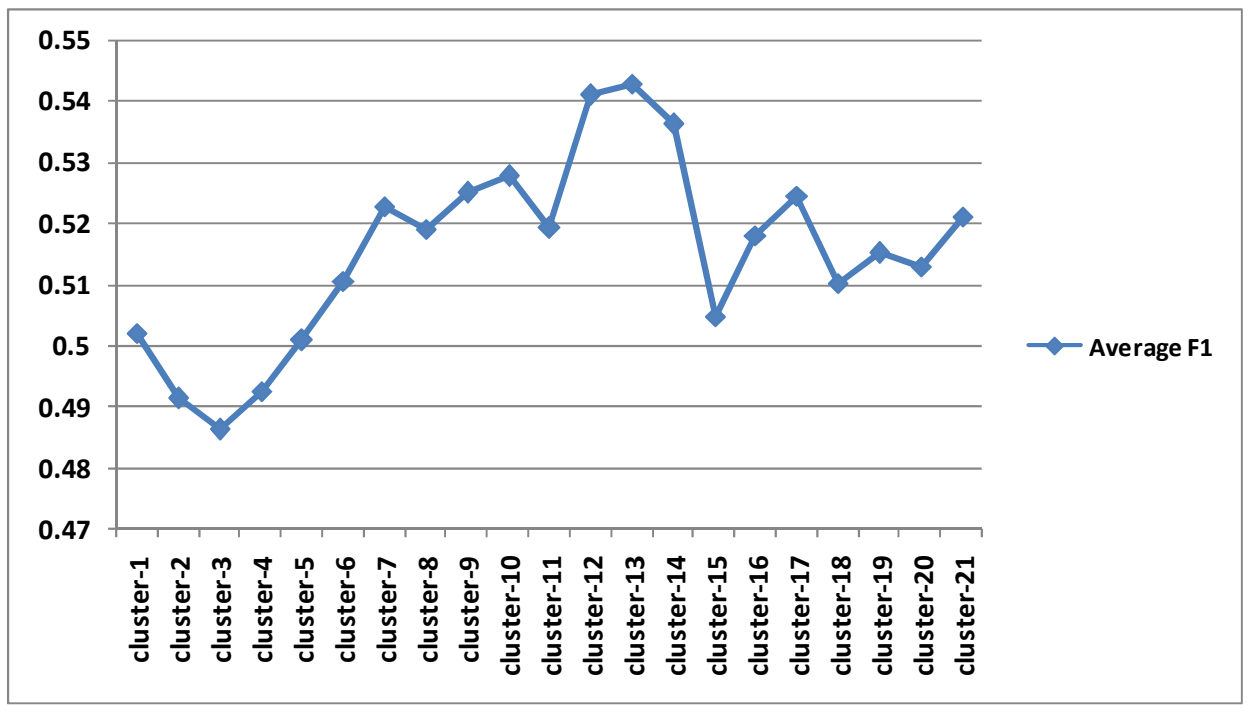

Figure 5. Average F1 measure diagram for each cluster in the contextual recommendation approach

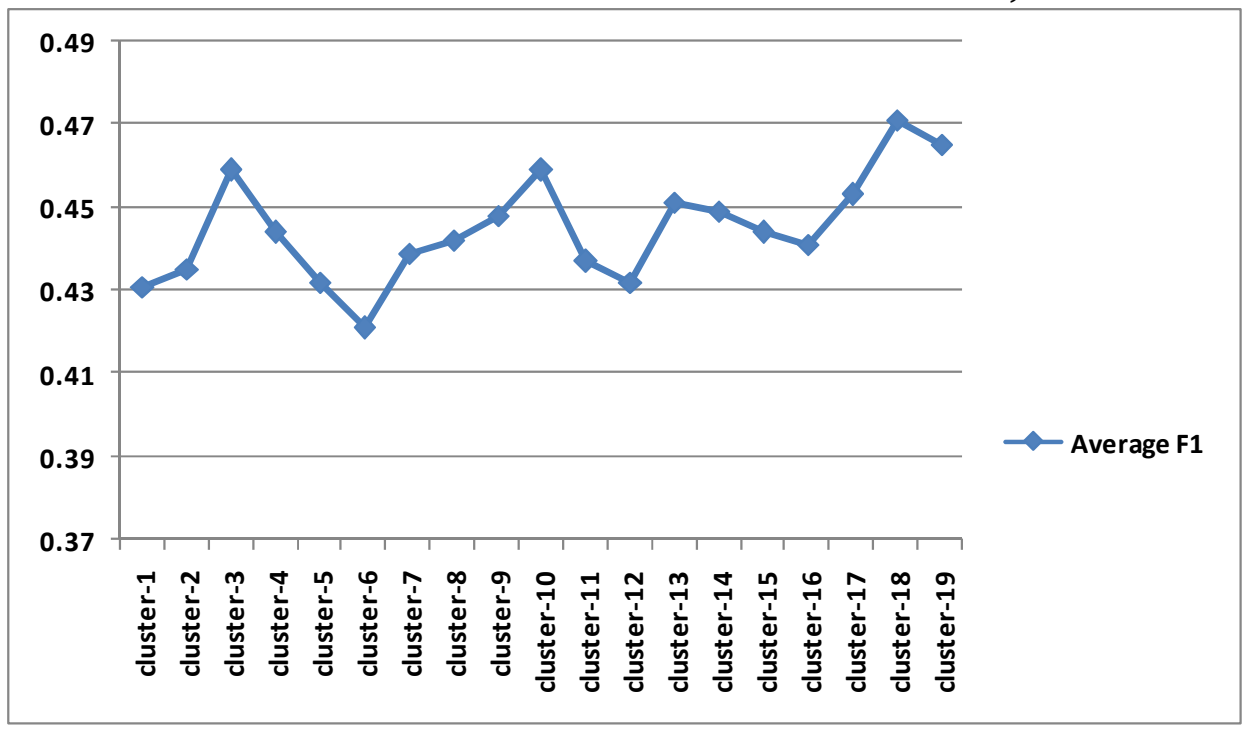

Figure 6. Average F1 measure diagram for each cluster in the contextual recommendation approach

## VI. CONCLUSION

In this paper a multidimensional approach for recommendation in m-commerce has been presented. The approach includes three phase, recognizing users' usage patterns, making a new 2-dimensional space and doing the final recommendation. It has been tested in a restaurant foods recommendation system. Context parameters of the system are Day, Time, Companion and whether. The evaluation results illustrate the proper quality of the approach and the necessity of context in mobile recommender systems.